\documentclass{aastex631}

\usepackage{amsmath}
\usepackage{rotating}
\usepackage{makecell}
\usepackage{multirow}
\usepackage{booktabs}

\begin{document}

\title{Detailed TESS-Based Light Curve Modeling and Fundamental Parameter Estimation of\\
27 W UMa-Type Contact Binaries}

\author[0000-0002-0196-9732]{Atila Poro}
\affiliation{LUX, Observatoire de Paris, CNRS, PSL, 61 Avenue de l'Observatoire, 75014 Paris, France}
\affiliation{Astronomy Department of the Raderon AI Lab., BC., V5C 0J3 Burnaby, Canada}
\email{atila.poro@obspm.fr, atilaporo@bsnp.info}

\author{Ahmad Sarostad}
\affiliation{Yazd Branch of the National Organization for Development of Exceptional Talents (SAMPAD), 89156 Yazd, Iran}
\affiliation{Yazd Desert Night Sky Astronomy Institute, 89156 Yazd, Iran}

\author{Mohammad Davoudi}
\affiliation{Yazd Branch of the National Organization for Development of Exceptional Talents (SAMPAD), 89156 Yazd, Iran}

\author{Mahsa Marami}
\affiliation{Yazd Branch of the National Organization for Development of Exceptional Talents (SAMPAD), 89156 Yazd, Iran}

\author{Motaharehalsadat Mohtaram}
\affiliation{Yazd Branch of the National Organization for Development of Exceptional Talents (SAMPAD), 89156 Yazd, Iran}

\author{Mohammadparsa Sharififard}
\affiliation{Yazd Branch of the National Organization for Development of Exceptional Talents (SAMPAD), 89156 Yazd, Iran}

\author{Diana Zadkhosh}
\affiliation{Yazd Branch of the National Organization for Development of Exceptional Talents (SAMPAD), 89156 Yazd, Iran}

\author{Aida Rakian}
\affiliation{Yazd Branch of the National Organization for Development of Exceptional Talents (SAMPAD), 89156 Yazd, Iran}

\author{Alireza Shiri Ahmadabadi}
\affiliation{Yazd Branch of the National Organization for Development of Exceptional Talents (SAMPAD), 89156 Yazd, Iran}

\author{AmirAli Akbarzadeh}
\affiliation{Yazd Branch of the National Organization for Development of Exceptional Talents (SAMPAD), 89156 Yazd, Iran}

\author{Amirali Ashrafpour}
\affiliation{Yazd Branch of the National Organization for Development of Exceptional Talents (SAMPAD), 89156 Yazd, Iran}

\author{Aram Ahmadi}
\affiliation{Yazd Branch of the National Organization for Development of Exceptional Talents (SAMPAD), 89156 Yazd, Iran}

\author{Arshida Amiri}
\affiliation{Yazd Branch of the National Organization for Development of Exceptional Talents (SAMPAD), 89156 Yazd, Iran}

\author{Arshida Goudarzi}
\affiliation{Yazd Branch of the National Organization for Development of Exceptional Talents (SAMPAD), 89156 Yazd, Iran}

\author{Atena Ahmadzadeh}
\affiliation{Yazd Branch of the National Organization for Development of Exceptional Talents (SAMPAD), 89156 Yazd, Iran}

\author{Ehsan Abbasi}
\affiliation{Yazd Branch of the National Organization for Development of Exceptional Talents (SAMPAD), 89156 Yazd, Iran}

\author{Helia Dolati}
\affiliation{Yazd Branch of the National Organization for Development of Exceptional Talents (SAMPAD), 89156 Yazd, Iran}

\author{Kimia Foroush Bastani}
\affiliation{Yazd Branch of the National Organization for Development of Exceptional Talents (SAMPAD), 89156 Yazd, Iran}

\author{Mahdis Derakhshan}
\affiliation{Yazd Branch of the National Organization for Development of Exceptional Talents (SAMPAD), 89156 Yazd, Iran}

\author{MehrAna Mehraban Pour}
\affiliation{Yazd Branch of the National Organization for Development of Exceptional Talents (SAMPAD), 89156 Yazd, Iran}

\author{Nazanin Karimi Shoushtari}
\affiliation{Yazd Branch of the National Organization for Development of Exceptional Talents (SAMPAD), 89156 Yazd, Iran}

\author{Parsa Arezoumand}
\affiliation{Yazd Branch of the National Organization for Development of Exceptional Talents (SAMPAD), 89156 Yazd, Iran}

\author{Reyhaneh Saki}
\affiliation{Yazd Branch of the National Organization for Development of Exceptional Talents (SAMPAD), 89156 Yazd, Iran}

\author{Romina Ozhand}
\affiliation{Yazd Branch of the National Organization for Development of Exceptional Talents (SAMPAD), 89156 Yazd, Iran}

\author{Sama Ghonche Sefidi}
\affiliation{Yazd Branch of the National Organization for Development of Exceptional Talents (SAMPAD), 89156 Yazd, Iran}

\author{Shayan Saeidzadeh}
\affiliation{Yazd Branch of the National Organization for Development of Exceptional Talents (SAMPAD), 89156 Yazd, Iran}

\author{Yeganeh Chaji}
\affiliation{Yazd Branch of the National Organization for Development of Exceptional Talents (SAMPAD), 89156 Yazd, Iran}

\author{Zahra Izadi}
\affiliation{Yazd Branch of the National Organization for Development of Exceptional Talents (SAMPAD), 89156 Yazd, Iran}

\author{Negar Sarlak}
\affiliation{Yazd Branch of the National Organization for Development of Exceptional Talents (SAMPAD), 89156 Yazd, Iran}

\author{Ayda Davoudi}
\affiliation{Yazd Branch of the National Organization for Development of Exceptional Talents (SAMPAD), 89156 Yazd, Iran}

\author{Danial Rahbarmah}
\affiliation{Yazd Branch of the National Organization for Development of Exceptional Talents (SAMPAD), 89156 Yazd, Iran}

\author{Hossein Dehghan}
\affiliation{Yazd Branch of the National Organization for Development of Exceptional Talents (SAMPAD), 89156 Yazd, Iran}

\author{Fahri Alicavus}
\affiliation{Çanakkale Onsekiz Mart University, Faculty of Science, Department of Physics, 17020, Çanakkale, Türkiye}
\affiliation{Çanakkale Onsekiz Mart University, Astrophysics Research Center and Ulupnar Observatory, 17020, Çanakkale, Türkiye}

\begin{abstract}
We performed a comprehensive analysis of 27 short period contact binary systems for which no light curve analysis had previously been reported. Photometric time-series data from the TESS mission were used in this analysis. The observational results were validated using additional TESS sectors, along with complementary photometric observations from the ASAS-SN survey. The photometric light curves of the 27 contact binary systems were analyzed with the BSN application. Model solutions were obtained through iterative fitting followed by MCMC-based refinement to derive reliable system parameters, and starspot configurations were incorporated for seven targets exhibiting O'Connell effect asymmetries. The absolute parameters of the target systems were derived using the empirical parameter relationship between orbital period and semi-major axis. Based on the results of the light curve solutions and the estimated absolute parameters, five of the analyzed systems are identified as A-subtype, while the remaining targets belong to the W-subtype. We analyzed a sample of 484 W UMa contact binaries to identify the physical and orbital parameters that most effectively distinguish A-subtype from W-subtype systems using t-statistics. Three targets exhibited extremely low mass ratios, and their orbital analysis confirmed that they are dynamically stable. The evolutionary states of the systems were examined, showing that lower-mass companions are generally more evolved, while more massive components remain less evolved. The positions of the systems and their stellar components were compared across four diagrams derived from empirical parameter relationship studies, showing good agreement with the linear fits.
\end{abstract}

\keywords{Close binary stars - Eclipsing binary stars - Fundamental parameters of stars}

%%%%%%%%%%%%%%%%% BODY OF PAPER %%%%%%%%%%%%%%%%%%
\section{Introduction}
Contact binary systems are characterized by both components filling and exceeding their Roche lobes. This configuration leads to the formation of a shared convective envelope in which mass, energy, and angular momentum are efficiently redistributed (\citealt{1971ARA&A...9..183P,1981ApJ...245..650M}). This physical connection enforces strong thermal coupling between the stars and results in short orbital periods, nearly equal surface temperatures, and continuous light variations with eclipse depths of similar amplitude. The geometric configuration of a contact binary is described by the degree of overcontact, commonly quantified by the fillout factor, which specifies the system's position between the inner and outer critical Roche equipotential surfaces (\citealt{1968ApJ...151.1123L,1971ApJ...166..605W,2005ApJ...629.1055Y}).

Light curve modeling is complicated by the interdependence of key system parameters, particularly the orbital inclination, mass ratio, fillout factor, and third-light contribution. Additional complications arise from magnetic activity and surface inhomogeneities such as starspots, which are common in contact binary systems and can distort the observed photometric behavior. Consequently, the reliability of the derived fundamental parameters depends on consistent modeling strategies and careful treatment of uncertainties.

W UMa systems are commonly divided into two subtypes, A-subtype and W-subtype, based on the relative temperatures and masses of their components. In A-subtype systems, the more massive star is also the hotter component, a configuration typically associated with earlier spectral types (A-F) and relatively longer orbital periods within the W UMa family (\citealt{1970VA.....12..217B}). In contrast, W-subtype systems are characterized by the counterintuitive situation in which the more massive star is cooler than its less massive companion (\citealt{1970VA.....12..217B}). These systems are usually composed of later-type stars (G-K spectral categories) and tend to have shorter orbital periods (\citealt{2000AJ....120..319R}). Their light curves often display asymmetries and temporal variations, which are commonly attributed to enhanced magnetic activity, the presence of starspots, or cyclic changes in energy redistribution within the common convective envelope (\citealt{1994A&A...288..529M}). Consequently, the A- and W-type classification provides a useful framework for interpreting the observed photometric behavior and fundamental physical properties of W UMa-type contact binary systems.

Automated light curve modeling techniques have been applied to large samples of contact binary systems in several studies (e.g., \citealt{2020ApJS..247...50S}, \citealt{2022ApJS..262...12K}, \citealt{2023MNRAS.525.4596D}, \citealt{2024AJ....167..192D}, \citealt{2024ApJS..271...32L}). These investigations aimed to determine key physical parameters, such as component temperatures ($T_{1,2}$), mass ratios ($q$), orbital inclinations ($i$), and fillout factors ($f$). The study by \cite{2025MNRAS.538.1427P} compared the parameter values they derived with those previously obtained from automated light curve analyses for the same sample of 18 contact binary systems. Significant differences were found in the main parameters, particularly in temperature ratios, orbital inclinations, and fillout factors between the two sets of results. According to \cite{2025MNRAS.538.1427P}, these differences likely stem from the characteristics of the photometric data used in the automated light curve modeling method. Many of these studies relied on single-band survey ground-based observations, including unfiltered photometry from the Catalina Sky Survey (CSS) converted to $V$-band magnitudes, or $V$-band measurements from the All-Sky Automated Survey for Supernovae (ASAS-SN; \citealt{2018MNRAS.477.3145J}). Substantial uncertainties and their scattered photometric data may have influenced the accuracy of the derived light curve parameters (\citealt{2025MNRAS.538.1427P}). Therefore, these methods still require improvements and should rely on more reliable, multiband data, particularly from space-based observations. As a result, the analysis of contact binary systems continues to depend on case-by-case modeling.

W UMa-type contact binary systems continue to attract significant attention, as precise observations and detailed analyses are essential for improving our understanding of their structure and evolutionary behavior. In this study, we present the first detailed photometric light curve analysis of 27 W UMa-type contact binary systems using TESS data. The aim is to constrain their physical and fundamental parameters through photometric light curve modeling. The paper is organized as follows: Section 2 describes the target systems and the dataset used, Section 3 presents the photometric light curve modeling, Section 4 details the determination of absolute parameters, and Section 5 provides the discussion and conclusions.

%%%%%%%%%%%%%%%%%%%%%%%%%%%%%%%%%%%%%%%%%%%%%%%
\vspace{0.6cm}
\section{Dataset}
We randomly selected 27 W UMa-type contact binary systems for this study. All of these targets have orbital periods shorter than 0.5 days, and none of them has been investigated in detail in previous studies. The temperatures reported for these systems in Gaia DR3 (\citealt{2023A&A...674A..33G}) range from 4300 to 7500 K, and their apparent magnitudes span from 11.9 to 15.2 mag. These binaries are identified as contact systems in several astronomical catalogs and databases, including ASAS-SN and the Variable Star Index (VSX\footnote{\url{https://vsx.aavso.org/}}). Basic information of the target systems is provided in Table \ref{Tab:systemsinfo}. This table includes equatorial coordinates, distances, and system temperatures from Gaia DR3. The $V$-band maximum magnitudes were adopted from the VSX database, while the orbital periods ($P_0$) and reference epochs ($t_0$) were taken from ASAS-SN.

We extracted space-based time-series photometry from TESS, which employs four wide-field cameras to observe each sector of the sky for roughly 27.4 days. We collected TESS observations in the broad “TESS:T” band (600–1000 nm) for each system. The names of the systems in the TESS Input Catalog (TIC), the sectors utilized for the light curve analysis, and their respective exposure times are listed in Table \ref{Tab:systemsinfo}. All data were downloaded from the Mikulski Archive for Space Telescopes (MAST) and processed using the Lightkurve package, applying detrending procedures aligned with the Science Processing Operations Center (SPOC) pipeline.

\begin{table*}
\renewcommand\arraystretch{1.2}
\caption{General information about the selected systems.}
\centering
\begin{center}
\footnotesize
\begin{tabular}{c c c c c c c c c c}
\hline
System & RA$.^\circ$ & Dec$.^\circ$ & $d$ & $T$ & $V_{\text{max}}$ & $P_0$ & $t_0$ & TESS & Exposure\\

TIC & (J2000) & (J2000) & (pc) & (K) & (mag) & (day) & ($\mathrm{BJD}_{\mathrm{TDB}}$) & sector & time(s)\\
\hline
5800141	&	79.0014438	&	22.6179799	&	244(1)	&	5386	&	11.93	&	0.3295210	&	2458074.99246	&	45	&	600	\\
23460412	&	108.3298035	&	-1.5880400	&	868(63)	&	5701*	&	14.48	&	0.3650174	&	2457475.56637	&	7	&	1800	\\
41740747	&	232.9134919	&	37.7329768	&	581(3)	&	5467	&	13.33	&	0.3075541	&	2457852.94265	&	77	&	200	\\
48219491	&	283.0303611	&	50.7538723	&	1133(15)	&	5969	&	14.19	&	0.3844300	&	2456801.02220	&	82	&	200	\\
80428640	&	10.9305862	&	-42.7735792	&	643(7)	&	5143	&	13.95	&	0.3091121	&	2457150.90878	&	29	&	600	\\
82710288	&	256.5321094	&	41.3083421	&	792(6)	&	5889	&	13.37	&	0.3620256	&	2457056.13556	&	79	&	200	\\
84445315	&	101.0977286	&	25.6435029	&	750(12)	&	5350	&	13.70	&	0.3413200	&	2456629.86728	&	72	&	200	\\
89667365	&	275.6525089	&	-46.7532421	&	766(10)	&	5988	&	13.18	&	0.3795577	&	2457078.86758	&	93	&	200	\\
101040463	&	56.5378121	&	-42.7404579	&	969(14)	&	5555	&	14.56	&	0.3081895	&	2458176.60281	&	4	&	1800	\\
138759051	&	61.2676767	&	-14.1682381	&	600(6)	&	5106	&	13.64	&	0.3262336	&	2458036.78496	&	31	&	600	\\
158546470	&	164.3466526	&	-39.0331112	&	783(9)	&	5804*	&	13.81	&	0.3406025	&	2456976.82681	&	90	&	200	\\
184565222	&	31.3029261	&	39.1735227	&	754(13)	&	5385	&	13.22	&	0.3663311	&	2457422.74796	&	85	&	200	\\
192854410	&	83.7417713	&	-38.4242886	&	286(1)	&	5253	&	12.86	&	0.2617414	&	2456927.80767	&	87	&	200	\\
231718985	&	80.5448014	&	-48.0158361	&	1284(27)	&	6166	&	14.32	&	0.3853035	&	2458087.74258	&	5	&	1800	\\
243212894	&	17.2406384	&	21.7776811	&	349(2)	&	5231	&	12.96	&	0.3080886	&	2456982.79267	&	57	&	200	\\
269943198	&	311.2070611	&	-26.7912125	&	808(12)	&	5531	&	13.73	&	0.3499972	&	2458213.87374	&	27	&	600	\\
276348408	&	228.2928352	&	-39.7959210	&	1366(36)	&	7461	&	12.82	&	0.4916847	&	2457534.76714	&	65	&	200	\\
280651790	&	144.5742862	&	-48.5092217	&	594(4)	&	5898	&	12.03	&	0.3679122	&	2458153.73374	&	62	&	200	\\
287602202	&	198.2593793	&	-45.5911398	&	799(13)	&	5573	&	12.20	&	0.3495288	&	2457485.80730	&	64	&	200	\\
293775345	&	98.3713483	&	-43.2336429	&	737(6)	&	6012	&	13.29	&	0.3427076	&	2457437.63795	&	88	&	200	\\
296861174	&	158.4134511	&	-14.2343683	&	685(10)	&	5780	&	13.89	&	0.2709752	&	2458126.06982	&	9	&	1800	\\
298708524	&	324.5215342	&	24.0465458	&	213(1)	&	4300	&	12.75	&	0.2397882	&	2457582.99849	&	82	&	200	\\
322580598	&	93.5393080	&	59.1018571	&	739(9)	&	5106	&	12.19	&	0.3115159	&	2457099.83577	&	73	&	200	\\
374271988	&	159.5591210	&	5.6753926	&	855(29)	&	5805	&	13.40	&	0.3215258	&	2457099.87922	&	72	&	200	\\
386768115	&	208.5051538	&	-17.7674999	&	1463(63)	&	5141	&	15.16	&	0.3563479	&	2457383.85734	&	91	&	120	\\
458718652	&	245.0467327	&	23.3361038	&	444(3)	&	5660	&	13.00	&	0.3138500	&	2458327.82574	&	79	&	200	\\
468265910	&	5.1037371	&	78.2410839	&	671(7)	&	5007	&	14.00	&	0.3438693	&	2457324.73380	&	85	&	200	\\
\hline
\end{tabular}
\end{center}
\vspace{2pt}
\begin{flushleft}
\footnotesize
\textit{*} The effective temperatures of two systems are not reported in Gaia DR3; the corresponding values reported in the table were taken from the TIC.
\end{flushleft}
\label{Tab:systemsinfo}
\end{table*}

%%%%%%%%%%%%%%%%%%%%%%%%%%%%%%%%%%%%%%%%%%%%%%%
\vspace{0.6cm}
\section{Light Curve Analysis}
BSN application (version 1.0; \citealt{2025Galax..13...74P}), developed specifically for modeling contact binary systems, was applied to the photometric light curve analysis of the targets. According to \citealt{2025Galax..13...74P}, the software offers an enhanced modeling environment that integrates an expanded suite of diagnostic tools, improved numerical stability, and a user-oriented workflow suitable for light curve solutions. The interface enables efficient parameter adjustment, clear visualization of results, and iterative refinement during the modeling process. The current version of the software is released for the Windows operating system.

At the outset of the photometric investigation, a uniform framework of physical and modeling assumptions was defined for all target systems. Contact geometry was adopted for every solution, since the shapes of the observed light curves, catalog identifications, and short orbital periods jointly imply physical and thermal contact between the components. The observational times, expressed in Barycentric Julian Date (BJD), were converted into orbital phase using the ephemerides listed in Table \ref{Tab:systemsinfo}. Gravity-darkening coefficients were fixed at \( g_{1}=g_{2}=0.32 \), following the prescription of \cite{1967ZA.....65...89L}. Bolometric albedos were set to \( A_{1}=A_{2}=0.5 \) in accordance with \cite{1969AcA....19..245R}. Stellar atmospheric properties for both components were represented using the model atmospheres of \cite{2004A&A...419..725C}. Within the BSN application environment, linear and logarithmic limb-darkening laws were applied, adopting coefficients from the tabulations of \cite{1993AJ....106.2096V}. Target system effective temperatures (\(T\)) were adopted from the Gaia DR3 catalog (Table \ref{Tab:systemsinfo}). Systems TIC 23460412 and TIC 158546470, lacking entries in Gaia DR3, had their temperature values taken from version 8.2 of the TESS Input Catalog. Initial temperatures reported in Gaia DR3 and TIC were assumed to correspond to the hotter components of each binary, as inferred from the relative depths of the light curve minima. Subsequently, the temperature of the cooler component was estimated based on the depth difference between the primary and secondary minima.

\begin{figure}
\centering
\includegraphics[width=0.7\textwidth]{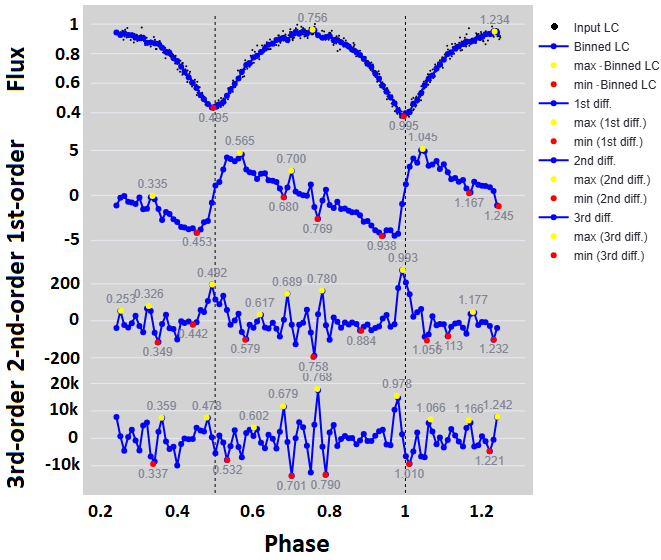}
\caption{The photometric light curve of TIC 184565222, together with its first through third derivatives (displayed from the top to the bottom panels), is presented as a sample of estimating the initial mass ratio for the targets. TIC 184565222 exhibits a mass ratio of \(q = 0.7714(102)\) using the Kouzuma method. It should be noted that in this method, the more massive component is introduced as \(M_1\) and the less massive component as \(M_2\). The vertical-axis units for these panels are W~m\(^{-2}\), 10~W~m\(^{-2}\)~day\(^{-1}\), \(10^{2}\)~W~m\(^{-2}\)~day\(^{-2}\), and \(10^{4}\)~W~m\(^{-2}\)~day\(^{-3}\), respectively.}
\label{fig:q-Kouzuma}
\end{figure}

Initial estimates of the mass ratio ($q$) were obtained using a novel approach to provide starting values for the light curve analysis. The method introduced by \cite{2023ApJ...958...84K, 2025PASJ...77.1323K} estimates the photometric mass ratio of overcontact binaries by analyzing higher-order derivatives of their light curves. The method has been previously discussed and evaluated in \cite{2024AJ....168..272P}. Implementation of the procedure involves calculating the third-order derivative, identifying local extrema near the eclipse phases, and combining these with the orbital period to define a parameter $W$, which correlates strongly with the mass ratio. Comparisons with systems having known spectroscopic mass ratios demonstrate that approximately 67\% of photometric estimates fall within the uncertainties, while 95\% deviate by less than $\pm0.1$. The method requires clearly defined maxima and minima in the relevant derivatives, as discussed by \cite{2023ApJ...958...84K, 2025PASJ...77.1323K}. This derivative-based technique was applied to all 27 target systems, providing initial mass ratios that served as the starting point for iterative light curve modeling, ultimately producing the final mass ratios for each system. Figure \ref{fig:q-Kouzuma} illustrates an example of the calculation process carried out for the TIC 184565222 system.

We used the TESS photometric data from the sectors listed in Table \ref{Tab:systemsinfo}, together with the initial parameter estimates, to construct theoretical fits to the light curves. The selected sectors correspond to the most recent available observations for each system, with preference given to datasets obtained with shorter exposure times whenever possible. Throughout the analysis procedure, all available additional sectors and the $V$-band photometric data from the ASAS-SN survey were consistently used for all systems to validate the synthetic light curves and the derived results.

Seven of the target systems display an asymmetry between the two light curve maxima, commonly known as the O'Connell effect. Accurate modeling of their observed light curves necessitated the inclusion of a starspot on one of the components (Table \ref{Tab:lc-analysis}). The TIC 23460412 system is unique in exhibiting a starspot on the hotter primary star, whereas all remaining systems show cool starspots located on the secondary component. This behavior is generally interpreted as a consequence of magnetic activity on the stellar surfaces, typically manifested through starspots, although alternative physical explanations have also been proposed in the literature (e.g., \citealt{1990ApJ...355..271Z,2003ChJAA...3..142L}).

The possible presence of third light ($l_3$) was investigated for all 27 systems during the light-curve analysis. In each case, $l_3$ was treated as a free parameter to test for any additional luminous contribution. Possible contamination from a nearby bright star within the photometric aperture was also examined as part of the modeling process. When multiple TESS sectors were available, independent light-curve solutions were derived for each sector and compared to assess the consistency of the results. In all systems, the fitted $l_3$ values were negligible within the uncertainties and their inclusion did not improve the fit quality. Consequently, third light was not included in the final models, indicating that the observed light curves of all systems are satisfactorily explained by binary configurations alone.

Application of the BSN optimization module was subsequently used to improve the parameter estimates, including the effective temperatures of both components ($T_{1,2}$), the mass ratio ($q$), the fillout factor ($f$), and the orbital inclination ($i$). Parameter determination, together with the associated uncertainties, was performed using a Markov Chain Monte Carlo (MCMC) approach. Efficient implementation of MCMC calculations within the BSN application framework enables rapid generation of synthetic light curves. The simulations adopted 24 walkers evolved over 2000 steps to explore the five principal parameters ($T_1$, $T_2$, $q$, $f$, and $i$). Removal of the initial 300 steps from each walker was carried out as a burn-in phase to ensure convergence of the chains. Statistical properties of the posterior distributions derived from the remaining samples were then used to extract the best-fitting values and their corresponding 1$\sigma$ uncertainties.

\begin{figure*}
\centering
\includegraphics[width=0.6\textwidth]{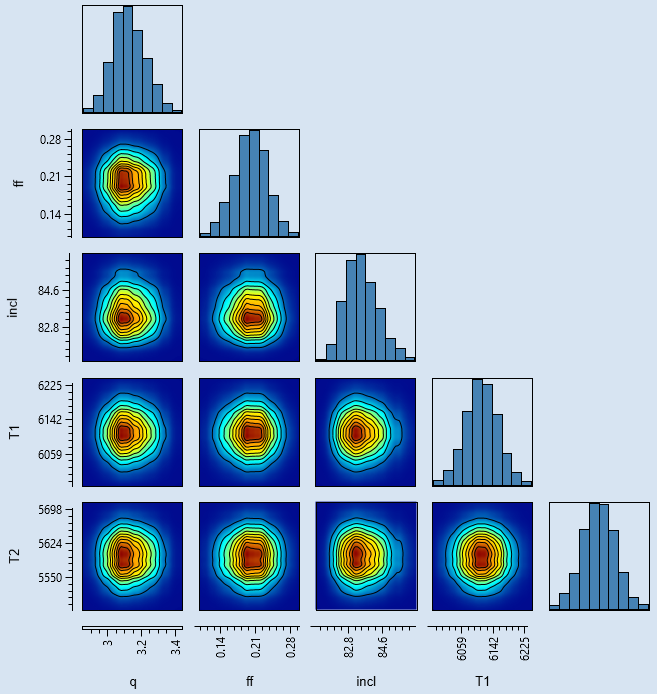}
\caption{Corner plot for the TIC 293775345 binary system, shown as a representative example among the target systems. The plot illustrates the posterior probability distributions and correlations among the five main parameters obtained from the MCMC analysis.}
\label{Fig:corner}
\end{figure*}

\begin{figure*}
\centering
\includegraphics[width=0.91\textwidth]{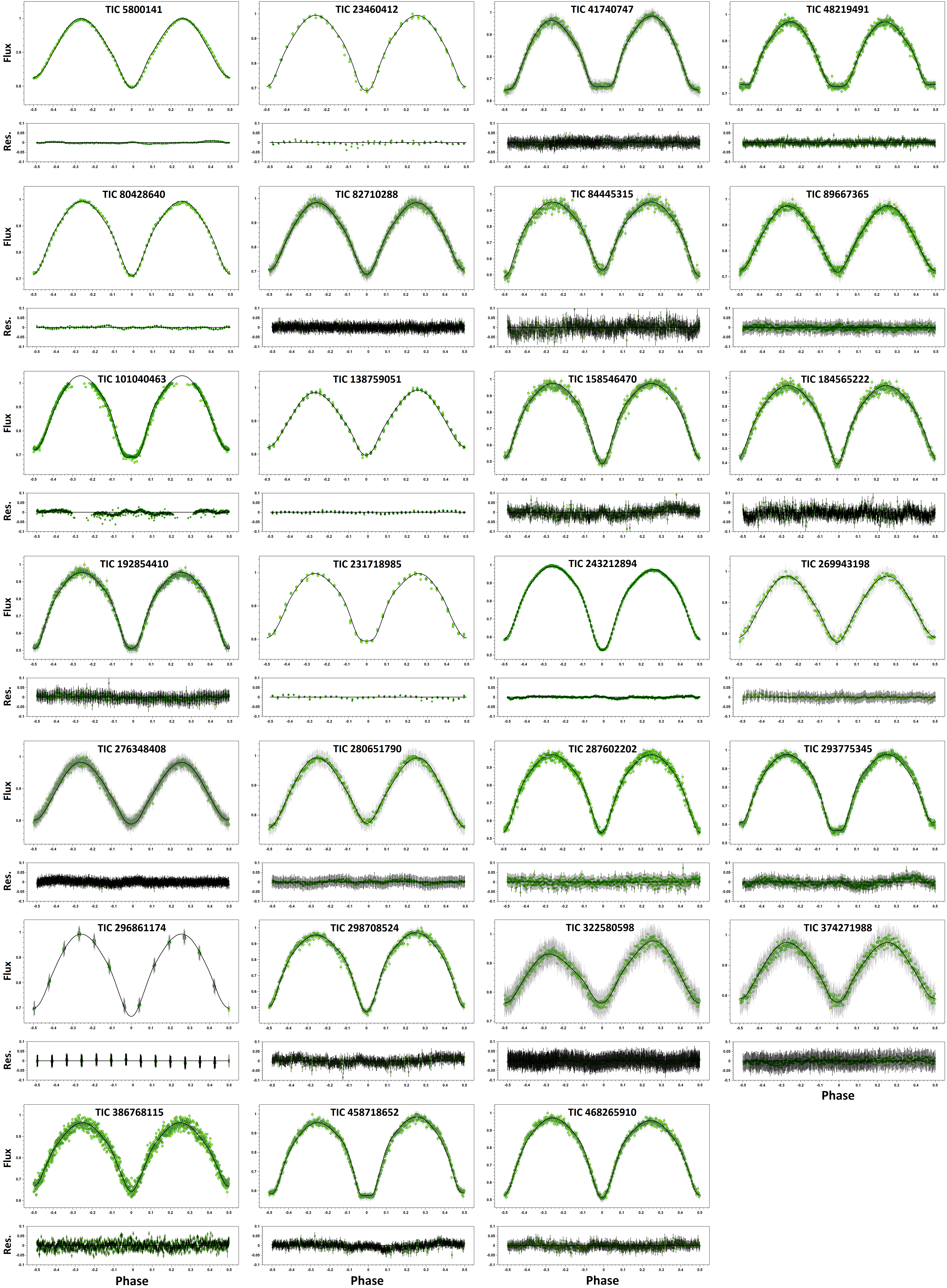}
\caption{Observed and synthetic light curves for the 27 contact binary systems. The residuals are shown in the bottom panel.}
\label{Fig:lc}
\end{figure*}

\begin{figure*}
\centering
\includegraphics[width=0.8\textwidth]{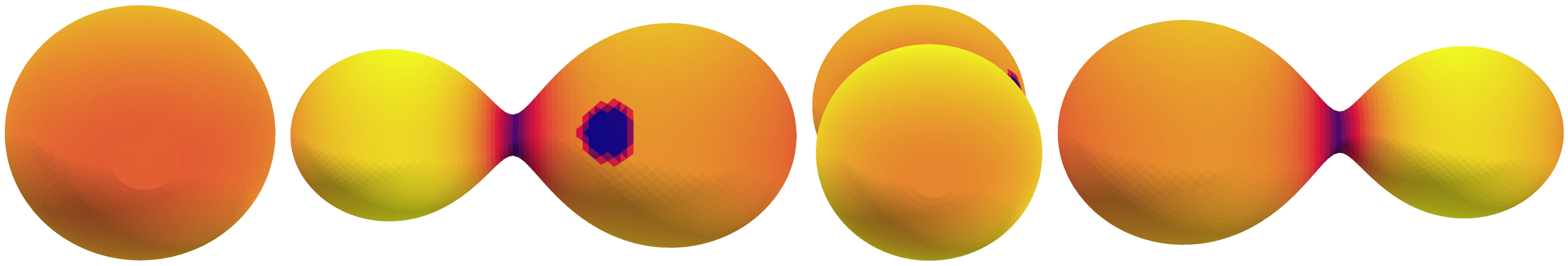}
\caption{Three-dimensional view of the TIC 468265910 binary system at four orbital phases (0, 0.25, 0.50, and 0.75, respectively). This system is shown as a representative example among the targets, illustrating the relative positions of the component stars throughout the orbit.}
\label{Fig:3d}
\end{figure*}

Posterior corner plots illustrating the parameter probability distributions and mutual correlations obtained from the MCMC analysis are presented for TIC 293775345 as a representative example in Figure \ref{Fig:corner}. Outcomes of the light curve modeling, including the inferred parameters together with their associated uncertainties, are listed in Table \ref{Tab:lc-analysis}. Final synthetic light curves overlaid on the TESS observed photometric data for all target binary systems are displayed in Figure \ref{Fig:lc}. Three-dimensional (3D) visual representations of the binary configurations are additionally shown in Figure \ref{Fig:3d}, where TIC 468265910 is adopted as a representative case.

\begin{sidewaystable*}
\renewcommand\arraystretch{1.3}
\setlength{\tabcolsep}{4pt}
\caption{Photometric light curve solutions for the target binary systems.}
\centering
\footnotesize
\begin{tabular}{c c c c c c c c c c c| c c c c c}
\hline
TIC & $T_{1}$ (K) & $T_{2}$ (K) & $q=M_2/M_1$ & $i^{\circ}$ & $f$ & $\Omega_1=\Omega_2$ & $l_1/l_{tot}$ & $l_2/l_{tot}$ & $r_{(mean)1}$ & $r_{(mean)2}$ & $Col.^\circ$ & $Long.^\circ$ & $Radius^\circ$ & $T_{spot}/T_{star}$ & Star\\
\hline
5800141	&	$5397_{\rm-(27)}^{+(31)}$	&	$4965_{\rm-(29)}^{+(31)}$	&	$4.294_{\rm-(3)}^{+(2)}$	&	$57.54_{\rm-(30)}^{+(22)}$	&	$0.229_{\rm-(2)}^{+(2)}$	&	8.138(131)	&	0.280(2)	&	0.720(3)	&	0.274(3)	&	0.522(5)	&	-	&	-	&	-	&	-	&	-	\\
23460412	&	$5714_{\rm-(40)}^{+(44)}$	&	$5532_{\rm-(38)}^{+(46)}$	&	$2.094_{\rm-(3)}^{+(2)}$	&	$68.06_{\rm-(70)}^{+(70)}$	&	$0.122_{\rm-(9)}^{+(5)}$	&	5.311(98)	&	0.369(3)	&	0.631(5)	&	0.326(4)	&	0.455(4)	&	-	&	-	&	-	&	-	&	-	\\
41740747	&	$5697_{\rm-(58)}^{+(94)}$	&	$5388_{\rm-(65)}^{+(99)}$	&	$6.274_{\rm-(98)}^{+(95)}$	&	$85.39_{\rm-(1.76)}^{+(1.15)}$	&	$0.759_{\rm-(65)}^{+(82)}$	&	10.227(145)	&	0.226(2)	&	0.774(6)	&	0.274(4)	&	0.573(7)	&	69(1)	&	306(2)	&	15(1)	&	1.36(4)	&	Primary	\\
48219491	&	$5862_{\rm-(31)}^{+(41)}$	&	$6183_{\rm-(92)}^{+(61)}$	&	$0.112_{\rm-(2)}^{+(5)}$	&	$77.73_{\rm-(1.70)}^{+(1.91)}$	&	$0.722_{\rm-(99)}^{+(52)}$	&	1.943(21)	&	0.832(5)	&	0.168(1)	&	0.595(7)	&	0.241(5)	&	-	&	-	&	-	&	-	&	-	\\
80428640	&	$5116_{\rm-(20)}^{+(22)}$	&	$5097_{\rm-(23)}^{+(19)}$	&	$1.474_{\rm-(3)}^{+(6)}$	&	$66.52_{\rm-(30)}^{+(50)}$	&	$0.041_{\rm-(2)}^{+(4)}$	&	4.464(76)	&	0.417(3)	&	0.583(3)	&	0.350(7)	&	0.418(7)	&	-	&	-	&	-	&	-	&	-	\\
82710288	&	$5932_{\rm-(82)}^{+(79)}$	&	$5825_{\rm-(70)}^{+(50)}$	&	$3.254_{\rm-(122)}^{+(192)}$	&	$68.66_{\rm-(71)}^{+(65)}$	&	$0.281_{\rm-(35)}^{+(47)}$	&	6.776(56)	&	0.278(2)	&	0.722(4)	&	0.299(5)	&	0.502(6)	&	-	&	-	&	-	&	-	&	-	\\
84445315	&	$5160_{\rm-(65)}^{+(59)}$	&	$5315_{\rm-(46)}^{+(56)}$	&	$1.752_{\rm-(4)}^{+(5)}$	&	$82.00_{\rm-(70)}^{+(60)}$	&	$0.202_{\rm-(8)}^{+(5)}$	&	4.778(88)	&	0.354(2)	&	0.646(3)	&	0.349(4)	&	0.446(4)	&	-	&	-	&	-	&	-	&	-	\\
89667365	&	$5955_{\rm-(99)}^{+(65)}$	&	$5982_{\rm-(79)}^{+(45)}$	&	$3.802_{\rm-(36)}^{+(42)}$	&	$65.57_{\rm-(79)}^{+(78)}$	&	$0.455_{\rm-(40)}^{+(37)}$	&	7.373(111)	&	0.243(1)	&	0.757(4)	&	0.298(5)	&	0.524(5)	&	-	&	-	&	-	&	-	&	-	\\
101040463	&	$5618_{\rm-(28)}^{+(48)}$	&	$5131_{\rm-(21)}^{+(36)}$	&	$5.565_{\rm-(38)}^{+(26)}$	&	$76.55_{\rm-(25)}^{+(49)}$	&	$0.409_{\rm-(74)}^{+(31)}$	&	9.595(87)	&	0.249(2)	&	0.751(3)	&	0.265(3)	&	0.550(7)	&	-	&	-	&	-	&	-	&	-	\\
138759051	&	$5223_{\rm-(22)}^{+(35)}$	&	$4883_{\rm-(45)}^{+(51)}$	&	$5.603_{\rm-(49)}^{+(48)}$	&	$58.04_{\rm-(27)}^{+(47)}$	&	$0.296_{\rm-(10)}^{+(14)}$	&	9.714(64)	&	0.229(1)	&	0.771(3)	&	0.257(5)	&	0.546(6)	&	120(1)	&	279(2)	&	12(1)	&	0.95(1)	&	Secondary	\\
158546470	&	$5918_{\rm-(50)}^{+(67)}$	&	$5723_{\rm-(52)}^{+(63)}$	&	$1.438_{\rm-(7)}^{+(8)}$	&	$81.99_{\rm-(50)}^{+(70)}$	&	$0.129_{\rm-(5)}^{+(6)}$	&	4.359(59)	&	0.450(3)	&	0.550(3)	&	0.360(6)	&	0.423(6)	&	-	&	-	&	-	&	-	&	-	\\
184565222	&	$5337_{\rm-(80)}^{+(63)}$	&	$5098_{\rm-(55)}^{+(79)}$	&	$1.199_{\rm-(6)}^{+(5)}$	&	$88.25_{\rm-(40)}^{+(40)}$	&	$0.242_{\rm-(6)}^{+(5)}$	&	3.931(61)	&	0.506(4)	&	0.494(4)	&	0.386(5)	&	0.419(5)	&	-	&	-	&	-	&	-	&	-	\\
192854410	&	$5377_{\rm-(52)}^{+(47)}$	&	$5202_{\rm-(78)}^{+(68)}$	&	$2.150_{\rm-(4)}^{+(4)}$	&	$87.90_{\rm-(60)}^{+(50)}$	&	$0.219_{\rm-(6)}^{+(5)}$	&	5.331(63)	&	0.369(4)	&	0.631(5)	&	0.332(4)	&	0.464(5)	&	-	&	-	&	-	&	-	&	-	\\
231718985	&	$6191_{\rm-(22)}^{+(40)}$	&	$5972_{\rm-(18)}^{+(30)}$	&	$9.178_{\rm-(4)}^{+(4)}$	&	$69.04_{\rm-(42)}^{+(60)}$	&	$0.167_{\rm-(2)}^{+(3)}$	&	14.033(149)	&	0.140(1)	&	0.860(5)	&	0.217(8)	&	0.579(8)	&	-	&	-	&	-	&	-	&	-	\\
243212894	&	$5369_{\rm-(25)}^{+(32)}$	&	$5116_{\rm-(27)}^{+(24)}$	&	$2.508_{\rm-(37)}^{+(18)}$	&	$81.54_{\rm-(25)}^{+(34)}$	&	$0.122_{\rm-(7)}^{+(8)}$	&	5.882(34)	&	0.352(2)	&	0.648(3)	&	0.311(5)	&	0.471(6)	&	126(1)	&	143(1)	&	25(1)	&	0.91(1)	&	Secondary	\\
269943198	&	$5485_{\rm-(44)}^{+(68)}$	&	$5367_{\rm-(59)}^{+(43)}$	&	$3.001_{\rm-(15)}^{+(12)}$	&	$59.97_{\rm-(45)}^{+(48)}$	&	$0.171_{\rm-(17)}^{+(22)}$	&	6.512(52)	&	0.292(2)	&	0.708(4)	&	0.300(5)	&	0.489(5)	&	-	&	-	&	-	&	-	&	-	\\
276348408	&	$7530_{\rm-(61)}^{+(64)}$	&	$7297_{\rm-(49)}^{+(47)}$	&	$6.216_{\rm-(6)}^{+(4)}$	&	$58.44_{\rm-(60)}^{+(60)}$	&	$0.740_{\rm-(16)}^{+(14)}$	&	10.170(89)	&	0.199(1)	&	0.801(5)	&	0.274(3)	&	0.571(5)	&	-	&	-	&	-	&	-	&	-	\\
280651790	&	$5681_{\rm-(105)}^{+(82)}$	&	$5993_{\rm-(124)}^{+(90)}$	&	$3.798_{\rm-(166)}^{+(107)}$	&	$63.65_{\rm-(1.68)}^{+(1.23)}$	&	$0.364_{\rm-(95)}^{+(97)}$	&	7.425(75)	&	0.209(1)	&	0.791(4)	&	0.293(4)	&	0.519(7)	&	-	&	-	&	-	&	-	&	-	\\
287602202	&	$5556_{\rm-(46)}^{+(50)}$	&	$5551_{\rm-(58)}^{+(64)}$	&	$1.622_{\rm-(4)}^{+(5)}$	&	$80.03_{\rm-(50)}^{+(70)}$	&	$0.111_{\rm-(8)}^{+(7)}$	&	4.642(43)	&	0.395(4)	&	0.605(5)	&	0.348(5)	&	0.432(6)	&	-	&	-	&	-	&	-	&	-	\\
293775345	&	$6104_{\rm-(41)}^{+(42)}$	&	$5589_{\rm-(37)}^{+(37)}$	&	$3.118_{\rm-(89)}^{+(110)}$	&	$83.44_{\rm-(79)}^{+(99)}$	&	$0.201_{\rm-(35)}^{+(32)}$	&	6.648(67)	&	0.323(5)	&	0.677(5)	&	0.296(4)	&	0.492(4)	&	-	&	-	&	-	&	-	&	-	\\
296861174	&	$5990_{\rm-(47)}^{+(51)}$	&	$5716_{\rm-(50)}^{+(74)}$	&	$3.937_{\rm-(6)}^{+(4)}$	&	$70.24_{\rm-(50)}^{+(40)}$	&	$0.465_{\rm-(6)}^{+(8)}$	&	7.539(80)	&	0.273(2)	&	0.727(4)	&	0.296(4)	&	0.527(7)	&	-	&	-	&	-	&	-	&	-	\\
298708524	&	$4547_{\rm-(88)}^{+(54)}$	&	$4395_{\rm-(76)}^{+(52)}$	&	$1.875_{\rm-(118)}^{+(88)}$	&	$85.75_{\rm-(1.15)}^{+(1.32)}$	&	$0.121_{\rm-(31)}^{+(32)}$	&	5.002(48)	&	0.406(6)	&	0.594(6)	&	0.335(6)	&	0.446(6)	&	94(1)	&	243(2)	&	20(1)	&	0.96(1)	&	Secondary	\\
322580598	&	$5067_{\rm-(122)}^{+(138)}$	&	$5102_{\rm-(93)}^{+(81)}$	&	$7.779_{\rm-(138)}^{+(148)}$	&	$64.39_{\rm-(1.19)}^{+(1.71)}$	&	$0.540_{\rm-(47)}^{+(59)}$	&	12.157(151)	&	0.153(1)	&	0.847(5)	&	0.245(4)	&	0.579(5)	&	82(1)	&	265(2)	&	16(1)	&	0.81(1)	&	Secondary	\\
374271988	&	$5845_{\rm-(39)}^{+(50)}$	&	$5742_{\rm-(45)}^{+(53)}$	&	$3.864_{\rm-(5)}^{+(5)}$	&	$60.41_{\rm-(80)}^{+(80)}$	&	$0.174_{\rm-(6)}^{+(9)}$	&	7.629(72)	&	0.245(2)	&	0.755(4)	&	0.280(5)	&	0.510(5)	&	-	&	-	&	-	&	-	&	-	\\
386768115	&	$5116_{\rm-(26)}^{+(26)}$	&	$4947_{\rm-(22)}^{+(28)}$	&	$0.957_{\rm-(3)}^{+(2)}$	&	$69.49_{\rm-(20)}^{+(30)}$	&	$0.033_{\rm-(2)}^{+(5)}$	&	3.663(30)	&	0.544(5)	&	0.456(5)	&	0.387(7)	&	0.380(7)	&	-	&	-	&	-	&	-	&	-	\\
458718652	&	$5795_{\rm-(57)}^{+(59)}$	&	$5462_{\rm-(47)}^{+(56)}$	&	$2.824_{\rm-(6)}^{+(5)}$	&	$87.97_{\rm-(50)}^{+(50)}$	&	$0.150_{\rm-(5)}^{+(4)}$	&	6.290(69)	&	0.336(3)	&	0.664(5)	&	0.303(6)	&	0.483(6)	&	95(1)	&	290(2)	&	16(1)	&	0.84(1)	&	Secondary	\\
468265910	&	$5083_{\rm-(51)}^{+(55)}$	&	$4960_{\rm-(61)}^{+(57)}$	&	$1.769_{\rm-(6)}^{+(6)}$	&	$81.49_{\rm-(70)}^{+(90)}$	&	$0.104_{\rm-(5)}^{+(6)}$	&	4.860(45)	&	0.402(5)	&	0.598(5)	&	0.340(5)	&	0.439(7)	&	98(1)	&	56(1)	&	12(1)	&	0.85(1)	&	Secondary	\\
\hline
\end{tabular}
\label{Tab:lc-analysis}
\end{sidewaystable*}

%%%%%%%%%%%%%%%%%%%%%%%%%%%%%%%%%%%%%%%%%%%%%%%
\vspace{0.6cm}
\section{Estimation of Fundamental Parameters}
The empirical relationships between the orbital period and physical parameters were systematically investigated by \cite{2025PASP..137k4203P} using a statistical sample of 483 contact binary systems. The strength and stability of these relationships were found to vary among mass, radius, luminosity, and semi-major axis for each component, indicating that not all empirical relations possess the same statistical reliability. Analysis of the correlation coefficients shows that some parameters exhibit substantial scatter and are therefore less suitable for accurate estimation of absolute parameters. In contrast, the relationship between the orbital period and the semi-major axis displays the strongest linear correlation. Therefore, the orbital period–semi-major axis relationship, characterized by a correlation coefficient of 0.90, was adopted as the most stable and reliable empirical relation for estimating absolute parameters in this study.

An updated empirical calibration linking the orbital period and the semi-major axis in contact binary systems has been introduced by \cite{2024PASP..136b4201P}. This relation was established based on a homogeneous sample of 414 contact binaries, all of which have orbital periods shorter than 0.7 days. The resulting linear form of the period-semi-major axis relation is given in Equation (\ref{p-a}). In this formulation, the semi-major axis $a$ is expressed in units of solar radii ($R_\odot$), while the orbital period $P$ is measured in days.

\begin{equation}\label{p-a}
a=(0.372_{\rm-0.114}^{+0.113})+(5.914_{\rm-0.298}^{+0.272})\times P
\end{equation}

Once the semi-major axis was determined, the individual stellar masses were calculated by combining the mass ratio obtained from the light curve modeling with Kepler's third law. The masses of the primary and secondary components are given by Equations \ref{eq:M1} and \ref{eq:M2}:

\begin{eqnarray}
M_1 = \frac{4 \pi^2 a^3}{G P^2 (1+q)} \label{eq:M1},\
M_2 = q \times M_1 \label{eq:M2}.
\end{eqnarray}

Here, $M_1$ and $M_2$ represent the masses of the primary and secondary stars, respectively, $a$ is the semi-major axis, $P$ is the orbital period, $q$ is the mass ratio, and $G$ is the gravitational constant.

The stellar radii ($R$) were determined using the mean fractional radii ($r_{\rm mean,1}$ and $r_{\rm mean,2}$) listed in Table \ref{Tab:lc-analysis}, according to the relation $R_{1,2} = a \times r_{\rm mean,1,2}$. Once the effective temperatures and radii were established, the stellar luminosities were calculated. The absolute bolometric magnitudes ($M_{\rm bol}$) of both components were then derived from their luminosities using the standard formula, as shown in Equation \ref{eq:Mbol}:

\begin{equation}\label{eq:Mbol}
M_{\rm bol, 1,2} = M_{{\rm bol}, \odot} - 2.5 \log \left(\frac{L_{1,2}}{L_\odot}\right).
\end{equation}

In this equation, the Sun’s absolute bolometric magnitude is adopted as $M_{{\rm bol}, \odot} = 4.73$ (\citealt{2010AJ....140.1158T}). The surface gravities ($g$) of the components were computed on a logarithmic scale from the derived masses and radii. In addition, the orbital angular momentum ($J_0$) was estimated using the total mass, mass ratio, and orbital period according to Equation \ref{eqJ0}, following the formulation of \cite{2006MNRAS.373.1483E}:

\begin{equation}\label{eqJ0}
J_0 = \frac{q}{(1+q)^2} \left(\frac{G^2}{2 \pi} M^5 P \right)^{1/3}.
\end{equation}

The resulting absolute parameters for the five contact binary systems are summarized in Table \ref{Tab:absolute}.

\begin{sidewaystable*}
\renewcommand\arraystretch{1.2}
\setlength{\tabcolsep}{4pt}
\centering
\caption{Derived physical parameters of the targets. The parameters are presented following the order of their computation.}
\begin{tabular}{lcccccccccccc}
\hline
Target & $a(R_\odot)$ & $M_1(M_\odot)$ & $M_2(M_\odot)$ & $R_1(R_\odot)$ & $R_2(R_\odot)$ & $L_1(L_\odot)$ & $L_2(L_\odot)$ & $M_{\mathrm{bol},1}(mag.)$ & $M_{\mathrm{bol},2}(mag.)$ & $\log g_1$(cgs) & $\log g_2$(cgs) & $\log J_0$(cgs) \\
\hline
5800141   & 2.321(207) & 0.292(085) & 1.254(368) & 0.636(064) & 1.211(121) & 0.309(074) & 0.804(192) & 6.004(233) & 4.967(233) & 4.296(028) & 4.369(029) & 51.400(245) \\

23460412  & 2.531(218) & 0.528(148) & 1.105(312) & 0.825(082) & 1.151(110) & 0.654(160) & 1.119(265) & 5.191(237) & 4.608(231) & 4.327(025) & 4.359(029) & 51.608(237) \\

41740747  & 2.191(201) & 0.205(062) & 1.287(414) & 0.600(065) & 1.255(132) & 0.342(101) & 1.197(356) & 5.894(280) & 4.535(283) & 4.193(026) & 4.350(034) & 51.239(284) \\

48219491  & 2.646(223) & 1.513(416) & 0.169(054) & 1.574(153) & 0.638(068) & 2.637(616) & 0.536(154) & 3.677(228) & 5.408(274) & 4.224(025) & 4.058(033) & 51.253(234) \\

80428640  & 2.200(202) & 0.605(182) & 0.891(272) & 0.770(087) & 0.920(101) & 0.366(095) & 0.515(130) & 5.821(251) & 5.451(244) & 4.446(021) & 4.461(025) & 51.560(257) \\

82710288  & 2.513(217) & 0.382(108) & 1.244(427) & 0.751(078) & 1.262(125) & 0.630(181) & 1.652(427) & 5.231(275) & 4.185(250) & 4.268(022) & 4.331(046) & 51.508(289) \\

84445315  & 2.391(211) & 0.572(165) & 1.002(293) & 0.834(084) & 1.066(104) & 0.445(120) & 0.818(206) & 5.610(260) & 4.948(244) & 4.353(027) & 4.383(030) & 51.591(248) \\

89667365  & 2.617(222) & 0.348(096) & 1.322(383) & 0.780(080) & 1.371(130) & 0.689(196) & 2.170(542) & 5.134(272) & 3.889(242) & 4.195(021) & 4.285(031) & 51.507(241) \\

101040463 & 2.195(201) & 0.228(069) & 1.266(391) & 0.582(061) & 1.207(128) & 0.304(077) & 0.910(228) & 6.024(244) & 4.832(243) & 4.266(028) & 4.377(030) & 51.281(267) \\

138759051 & 2.301(206) & 0.233(068) & 1.305(398) & 0.591(066) & 1.257(128) & 0.235(061) & 0.809(212) & 6.304(252) & 4.960(253) & 4.261(021) & 4.355(032) & 51.311(265) \\

158546470 & 2.386(211) & 0.645(186) & 0.928(274) & 0.859(091) & 1.009(105) & 0.816(223) & 0.985(264) & 4.951(263) & 4.746(258) & 4.379(022) & 4.397(027) & 51.611(247) \\

184565222 & 2.538(218) & 0.744(209) & 0.892(256) & 0.980(098) & 1.064(105) & 0.702(194) & 0.689(187) & 5.114(265) & 5.135(261) & 4.327(025) & 4.335(028) & 51.668(234) \\

192854410 & 1.920(188) & 0.440(142) & 0.947(309) & 0.637(071) & 0.891(098) & 0.306(086) & 0.524(158) & 6.015(269) & 5.432(287) & 4.473(030) & 4.514(032) & 51.428(287) \\

231718985 & 2.651(223) & 0.165(045) & 1.519(418) & 0.575(071) & 1.535(152) & 0.438(127) & 2.701(615) & 5.626(276) & 3.651(223) & 4.137(004) & 4.247(023) & 51.247(226) \\

243212894 & 2.194(201) & 0.426(128) & 1.068(337) & 0.682(075) & 1.033(109) & 0.349(090) & 0.659(163) & 5.874(249) & 5.182(240) & 4.399(024) & 4.438(032) & 51.478(274) \\

269943198 & 2.442(213) & 0.399(114) & 1.197(349) & 0.733(077) & 1.194(118) & 0.438(119) & 1.066(270) & 5.627(262) & 4.660(245) & 4.309(022) & 4.362(030) & 51.521(241) \\

276348408 & 3.280(254) & 0.272(068) & 1.688(424) & 0.899(080) & 1.873(162) & 2.341(530) & 8.964(1904)& 3.807(222) & 2.349(209) & 3.964(023) & 4.120(025) & 51.529(201) \\

280651790 & 2.548(218) & 0.342(096) & 1.298(421) & 0.747(075) & 1.322(133) & 0.523(153) & 2.033(609) & 5.433(279) & 3.960(285) & 4.226(024) & 4.309(039) & 51.481(278) \\

287602202 & 2.439(213) & 0.608(174) & 0.986(286) & 0.849(087) & 1.054(108) & 0.619(160) & 0.950(256) & 5.251(250) & 4.785(259) & 4.364(024) & 4.386(026) & 51.614(245) \\

293775345 & 2.399(211) & 0.383(110) & 1.195(393) & 0.710(073) & 1.180(114) & 0.631(158) & 1.225(288) & 5.230(242) & 4.510(229) & 4.319(025) & 4.371(043) & 51.490(276) \\

296861174 & 1.975(191) & 0.285(091) & 1.122(360) & 0.584(065) & 1.041(116) & 0.396(109) & 1.042(301) & 5.735(265) & 4.685(276) & 4.359(028) & 4.453(029) & 51.316(281) \\

298708524 & 1.790(182) & 0.466(157) & 0.874(358) & 0.600(073) & 0.798(093) & 0.139(047) & 0.214(069) & 6.876(316) & 6.402(302) & 4.550(027) & 4.575(054) & 51.386(366) \\

322580598 & 2.214(202) & 0.171(051) & 1.331(431) & 0.543(059) & 1.282(129) & 0.175(063) & 1.004(298) & 6.623(335) & 4.726(282) & 4.202(024) & 4.346(038) & 51.178(281) \\

374271988 & 2.274(205) & 0.314(093) & 1.212(361) & 0.637(070) & 1.159(117) & 0.426(115) & 1.317(335) & 5.656(259) & 4.431(246) & 4.327(022) & 4.393(030) & 51.414(249) \\

386768115 & 2.479(215) & 0.823(233) & 0.788(227) & 0.960(102) & 0.942(101) & 0.569(142) & 0.479(120) & 5.343(242) & 5.529(242) & 4.389(021) & 4.386(022) & 51.651(243) \\

458718652 & 2.228(203) & 0.394(118) & 1.113(336) & 0.675(076) & 1.076(113) & 0.463(134) & 0.929(248) & 5.565(275) & 4.810(257) & 4.375(021) & 4.421(028) & 51.471(254) \\

468265910 & 2.406(212) & 0.571(164) & 1.010(295) & 0.818(085) & 1.056(111) & 0.403(109) & 0.609(171) & 5.718(260) & 5.269(269) & 4.369(024) & 4.395(024) & 51.597(244) \\
\hline
\end{tabular}
\label{Tab:absolute}
\end{sidewaystable*}

%%%%%%%%%%%%%%%%%%%%%%%%%%%%%%%%%%%%%%%%%%%%%%%%%%
\vspace{0.6cm}
\section{Discussion and Conclusion}
In this work, we carry out a comprehensive photometric investigation of 27 contact binary systems, deriving their absolute physical parameters from modeled light curves. The results obtained in this analysis form the basis for the interpretation of the systems' properties and support the discussion and conclusions presented in the following parts:

A) Light curve modeling results indicate that the effective temperatures of the stellar components in the analyzed contact binary systems span a range from 4395~K to 7530~K. Contact binaries are expected to exhibit nearly equal surface temperatures due to energy transfer through a common envelope, yet measurable temperature differences are still observed. In the 27 systems analyzed in this study, the temperature difference, $\Delta T = |T_1 - T_2|$, was calculated, with associated uncertainties obtained by propagating the individual temperature errors through quadrature addition (Table \ref{Tab:conclusion}). The minimum temperature difference occurs in TIC~287602202, with $\Delta T = 5$~K, indicating an almost complete thermal equilibrium, whereas the maximum contrast of $\Delta T = 515$~K is measured in TIC~293775345.  

The relative temperature difference can be expressed as a percentage:
\begin{equation}
\mathrm{DT\%} = \frac{|T_1 - T_2|}{T_1} \times 100.
\end{equation}

A comprehensive statistical investigation of relative temperature differences in contact binary systems has been carried out by \cite{2025PASP..137k4203P} using a sample of 763 objects. In that study, 132 systems were found to have DT\% below 1\%, about 251 systems exhibit DT\% below 2\%, the majority of the sample (504 systems) shows DT\% below 5\%, and nearly 696 systems ($\approx 91\%$) fall below the 10\% threshold, indicating that close thermal equilibrium is a common property among contact binaries. Only a small fraction, roughly 36 objects, display DT\% values exceeding 15\%, corresponding to cases with significant thermal imbalance. Comparison with this statistical framework shows that the temperature differences derived for the 27 target systems, fall within the expected range for contact binaries, with all systems exhibiting DT\% values below 10\%.

B) In contact binary systems, the fillout factor is used to quantify how strongly the stellar components overflow their Roche lobes. This parameter is defined as
\begin{equation}
\begin{aligned}
f = \frac{\Omega - \Omega_{\mathrm{in}}}{\Omega_{\mathrm{out}} - \Omega_{\mathrm{in}}},
\end{aligned}
\end{equation}
where $\Omega$ denotes the surface potential of the common envelope, and $\Omega_{\mathrm{in}}$ and $\Omega_{\mathrm{out}}$ represent the inner and outer critical Roche potentials, respectively. The fillout factor thus provides a convenient measure of the degree of contact and the geometrical configuration of the system.

Fillout factor classification was subsequently performed for targets using parameters derived from the light curve solutions. Based on established criteria by \cite{2022AJ....164..202L}, contact binary systems are categorized as deep-contact systems when $f \geq 50\%$, medium-contact systems when $25\% \leq f < 50\%$, and shallow-contact systems when $f < 25\%$. Therefor, 17 systems are identified as shallow-contact binaries, six targets belong to the medium-contact group, and four systems are classified as deep-contact objects; individual classifications are listed in Table \ref{Tab:conclusion}.

C) W UMa-type contact binaries are conventionally divided into two principal subclasses, namely A-type and W-type systems. Evolutionary studies have shown that these subtype follow distinct evolutionary paths (\citealt{2020MNRAS.492.4112Z}). Characteristic differences distinguish the two groups, such that the more massive component is hotter in A-type binaries, whereas the more massive star is cooler in W-type systems (\citealt{1970VA.....12..217B}). According to the present analysis, five targets are assigned to the A-subtype, while the remaining systems are classified as W-type, and subtype determinations for 27 systems are reported in Table \ref{Tab:conclusion}.

Additionally, previous studies have indicated that starspot activity occurs more frequently in W-subtype contact binary systems than in A-subtype systems (e.g., \citealt{2019PASJ...71...21K}). Seven systems required starspot modeling; only TIC 322580598 is classified as an A-subtype system, while the remaining six systems are W-subtype, which is consistent with earlier findings.

\begin{table*}
\renewcommand\arraystretch{1.2}
\caption{Temperature differences of the binary components, fillout factor classification, and subtype identification for each system derived from the results of this study.}
\centering
\begin{center}
\footnotesize
\begin{tabular}{c c c c| c c c c| c c c c}
\hline
TIC & $\Delta T$(K) & $f$ cat. & Subtype & TIC & $\Delta T$(K) & $f$ cat. & Subtype & TIC & $\Delta T$(K) & $f$ cat. & Subtype\\
\hline
5800141	&	432(42)	&	Shallow	&	W	&	138759051	&	340(56)	&	Medium	&	W	&	287602202	&	5(78)	&	Shallow	&	W	\\
23460412	&	182(59)	&	Shallow	&	W	&	158546470	&	195(83)	&	Shallow	&	W	&	293775345	&	515(56)	&	Shallow	&	W	\\
41740747	&	309(112)	&	Deep	&	W	&	184565222	&	239(98)	&	Shallow	&	W	&	296861174	&	274(79)	&	Medium	&	W	\\
48219491	&	321(85)	&	Deep	&	W	&	192854410	&	175(89)	&	Shallow	&	W	&	298708524	&	152(96)	&	Shallow	&	W	\\
80428640	&	19(30)	&	Shallow	&	W	&	231718985	&	219(39)	&	Shallow	&	W	&	322580598	&	35(156)	&	Deep	&	A	\\
82710288	&	107(101)	&	Medium	&	W	&	243212894	&	253(39)	&	Shallow	&	W	&	374271988	&	103(67)	&	Shallow	&	W	\\
84445315	&	155(80)	&	Shallow	&	A	&	269943198	&	118(76)	&	Shallow	&	W	&	386768115	&	169(36)	&	Shallow	&	A	\\
89667365	&	27(103)	&	Medium	&	A	&	276348408	&	233(79)	&	Deep	&	W	&	458718652	&	333(78)	&	Shallow	&	W	\\
101040463	&	487(48)	&	Medium	&	W	&	280651790	&	312(142)	&	Medium	&	A	&	468265910	&	123(79)	&	Shallow	&	W	\\
\hline
\end{tabular}
\end{center}
\label{Tab:conclusion}
\end{table*}

The distinction between A and W subtypes raises an important question: which physical and orbital parameters best separate these groups? A sample of 484 binaries from \cite{2025MNRAS.538.1427P} was analyzed to address this question. The parameters examined included the orbital period ($P$), mass ratio ($q$), inclination angle ($i$), fillout factor ($f$), effective temperatures of the primary and secondary components ($T_1$, $T_2$), masses ($M_1$, $M_2$), radii ($R_1$, $R_2$), luminosities ($L_1$, $L_2$), total system mass ($M_\mathrm{tot}$), temperature difference between components ($\Delta T$), radius ratio ($R_r = R_2/R_1$), and luminosity ratio ($L_r = L_2/L_1$). The sample was restricted to contact binary systems with orbital periods shorter than 0.6 days, ensuring consistency with the typical period range of W UMa binaries. Additionally, only systems with complete measurements for all examined parameters were included, resulting in a homogeneous dataset suitable for a robust statistical comparison between the two subtypes. It should be noted that the more massive star was considered as the primary component, and the remaining parameters were treated accordingly throughout the sample.

An independent two-sample t-test was applied to each parameter. This statistical method compares the means of two independent groups while accounting for sample variability. The t-statistic for a parameter $X$ is defined as

\begin{equation} \label{eq:tstat}
t = \frac{\bar{X}_A - \bar{X}_W}{\sqrt{\frac{s_A^2}{n_A} + \frac{s_W^2}{n_W}}}
\end{equation}
where $\bar{X}_A$ and $\bar{X}_W$ are the sample means of A-subtype and W-subtype binaries, $s_A^2$ and $s_W^2$ are the corresponding sample variances, and $n_A$ and $n_W$ represent the number of systems in each subtype. The absolute value of the t-statistic, $|t|$, serves as a direct measure of the parameter's discriminative power. Higher values indicate stronger separation between the two subtypes.

\begin{figure*}
\centering
\includegraphics[width=0.8\textwidth]{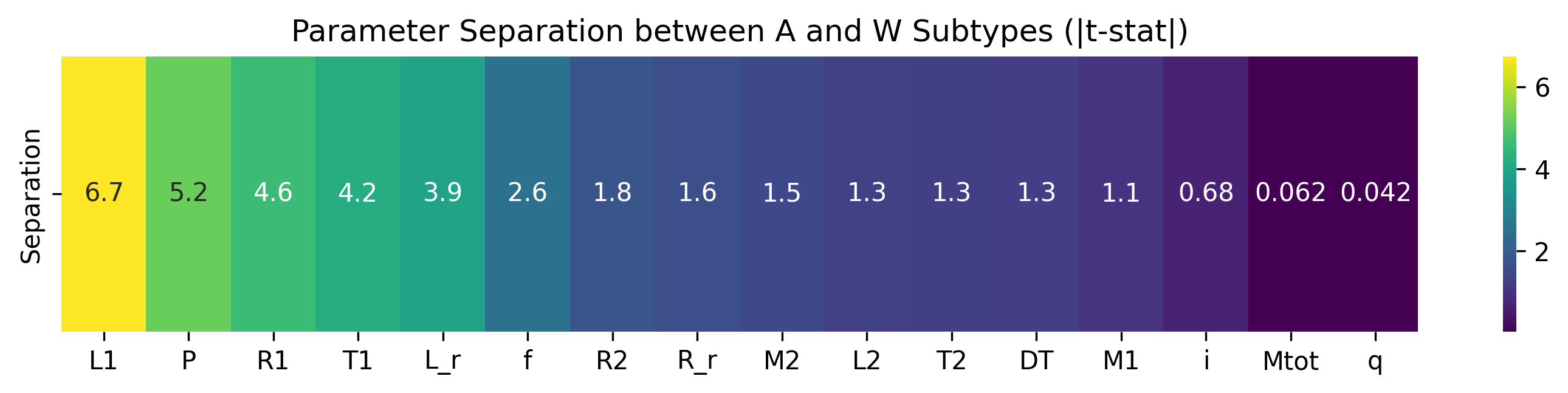}
\caption{Absolute t-statistics ($|t|$) for the separation of A-subtype and W-subtype W UMa binaries, indicating the discriminative power of each physical and orbital parameter.}
\label{Fig:AW}
\end{figure*}

Analysis of the sample systems revealed that the luminosity of the primary component ($L_1$) exhibited the most pronounced separation ($|t| \approx 6.74$). The orbital period ($P$, $|t| \approx 5.22$), the radius of the primary star ($R_1$, $|t| \approx 4.62$), and its effective temperature ($T_1$, $|t| \approx 4.20$) also showed strong differences. These results indicate that both the intrinsic brightness and physical size of the primary component, as well as the system's period, dominate the distinction between A-subtype and W-subtype binaries. Moderate contributions came from the relative luminosity ratio ($L_r$, $|t| \approx 3.93$) and the fillout factor ($f$, $|t| \approx 2.55$). Parameters of the secondary component, including $M_2$, $R_2$, $L_2$, and $T_2$, as well as inclination ($i$), total mass ($M_\mathrm{tot}$), and mass ratio ($q$), showed low separation values ($<1.8$), indicating minimal discriminative power. The discriminative power of each physical and orbital parameter between A-subtype and W-subtype binaries is illustrated in Figure \ref{Fig:AW}, where $|t|$-statistics are presented.

These findings suggest that the distinction between A-subtype and W-subtype contact binaries is primarily governed by the properties of the primary star and the system's orbital period. In general, A-subtype systems tend to have longer orbital periods, larger and more luminous primary stars, and higher primary temperatures compared to W-subtype systems. W-subtype binaries usually exhibit shorter periods, smaller and slightly cooler primary components, and correspondingly lower luminosities. The secondary star characteristics, total system mass, and mass ratio show minimal variation between subtypes and contribute little to differentiating A-subtype from W-subtype binaries. Overall, the dominant factors for subtype classification are the size, brightness, and temperature of the primary component along with the orbital period, providing a clear descriptive distinction between the two groups.

Analysis of the 27 contact binary systems shows that A-subtype systems tend to have longer orbital periods than W-subtype systems, with mean periods of 0.351 days and 0.335 days, respectively. Approximately 80\% of A-subtype systems exhibit periods longer than the average W-subtype period, indicating that the orbital period generally aligns with the conventional subtype classification. Temperature differences ($\Delta T$) between components reveal that A-subtype systems have an average contrast of about 140 K, while W-subtype systems display a larger difference of roughly 219 K, suggesting greater thermal disparity in W-subtype target binaries (Table \ref{Tab:conclusion}). Review of component masses indicates that A-subtype systems contain more massive stars averaging around $1.16~M_\odot$ and less massive stars around $0.44~M_\odot$. W-subtype systems feature slightly heavier primary components ($\sim 1.19~M_\odot$) and slightly lighter secondary components ($\sim 0.40~M_\odot$), showing differences in mass distribution between the two subtypes.

\begin{figure*}
\centering
\includegraphics[width=0.9\textwidth]{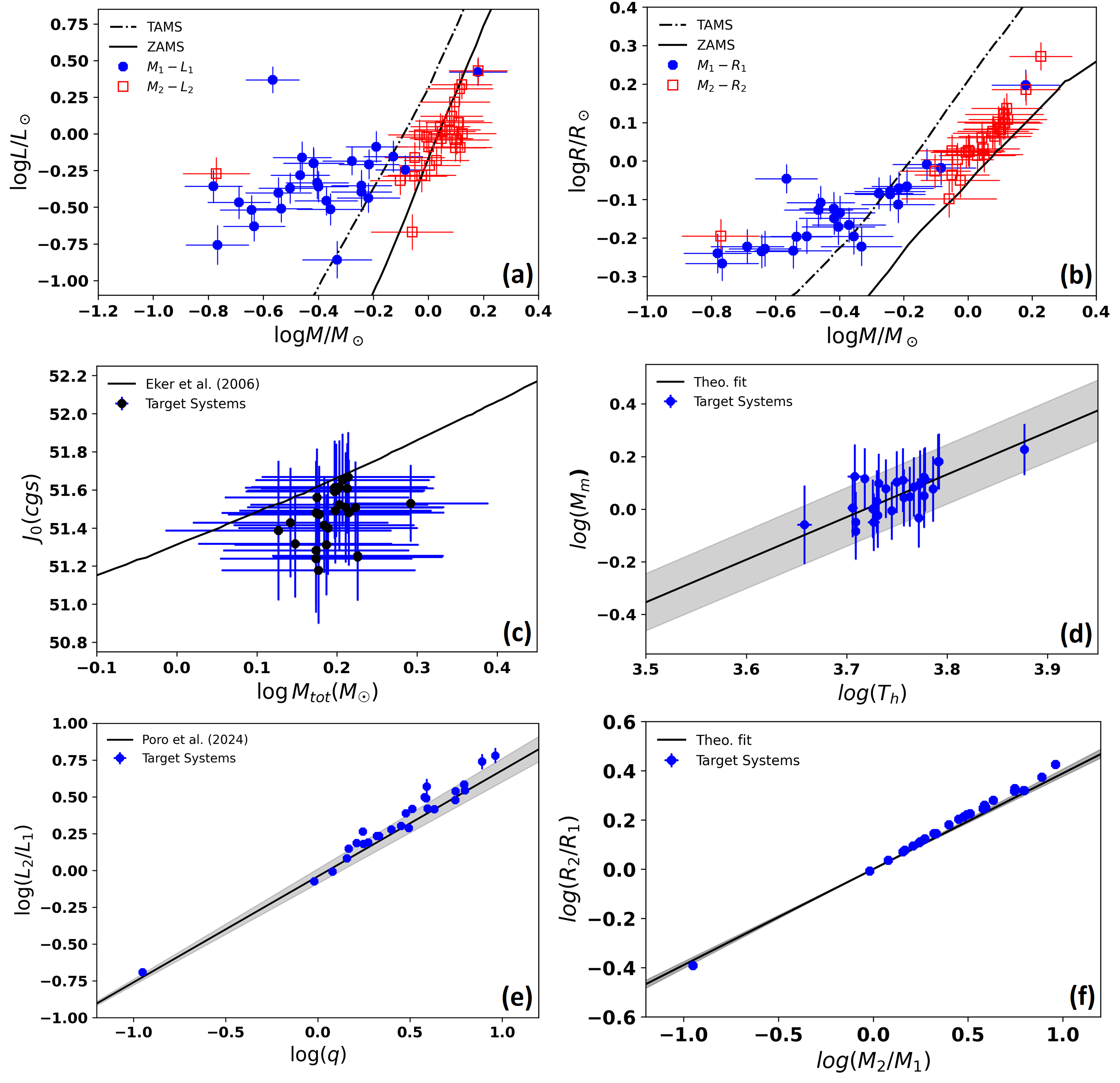}
\caption{Panels (a) and (b) show the $M-L$ and $M-R$ relationships relative to ZAMS and TAMS; panel (c) presents $M_\mathrm{tot}-J_0$ with the reference quadratic boundary; panel (d) displays the $T_h-M_m$ relationship; panels (e) and (f) illustrate the $L_\mathrm{ratio}-q$ and $R_\mathrm{ratio}-q$ relationships and are compared with empirical relationships.}
\label{Fig:rel}
\end{figure*}

D) Using derived absolute parameters, the Mass–Radius ($M$–$R$) and Mass–Luminosity ($M$–$L$) relations displayed on logarithmic scales are employed to evaluate the evolutionary state of the target systems (Figure \ref{Fig:rel}a,b). Relative positions of the stellar components are defined with respect to the Zero-Age Main Sequence (ZAMS) and Terminal-Age Main Sequence (TAMS) reference boundaries adopted from \cite{2000A&AS..141..371G}. As shown in Figure \ref{Fig:rel}a,b, lower-mass companions are found closer to or above the TAMS, whereas more massive stars tend to lie nearer to the ZAMS. Contact binary systems follow complex evolutionary pathways driven by continuous mass and angular momentum exchange (\citealt{2005ApJ...629.1055Y}), leading to substantial deviations from single-star evolutionary tracks. Consequently, any interpretation based on single-star ZAMS and TAMS boundaries should be treated with caution.

Orbital angular momentum values were derived for all systems and are listed in Table \ref{Tab:absolute}, while their positions are presented on the $J_{0}$-$\log M_{\mathrm{tot}}$ diagram (Figure \ref{Fig:rel}c). A quadratic reference relation adopted from \cite{2006MNRAS.373.1483E} is overplotted, indicating that the studied systems lie systematically below this boundary, within the parameter space typically associated with contact binary systems.

The study by \cite{2024RAA....24e5001P} presents the temperature–mass ($T_h$–$M_m$) relationship for contact binary systems. Positions of the systems are shown in Figure \ref{Fig:rel}d, compared to the linear relation fitted by \cite{2024RAA....24e5001P}:

\begin{equation}\label{Mm}
\log M_{m} = (1.6185 \pm 0.0150) \times (\log T_{h}) + (-6.0186 \pm 0.0562).
\end{equation}

Hotter stars and more massive component of each target system occupy positions on the $T_h$–$M_m$ diagram that align with the distribution of contact binaries reported by \cite{2024RAA....24e5001P}.

Figure \ref{Fig:rel}e,f presents the luminosity and radius ratios versus the mass ratio. These relationships were investigated by \cite{2024RAA....24a5002P} and \cite{2022PASP..134f4201P}, respectively. These studies reported $L_{\rm ratio}$–$q$ and $R_{\rm ratio}$–$q$ relations using linear fits (Equations \ref{eq:L-q}) and \ref{eq:R-q}.

\begin{equation}\label{eq:L-q}
\log (\frac{L_2}{L_1}) = (0.72 \pm 0.01)log q + (-0.04 \pm 0.02)
\end{equation}

\begin{equation}\label{eq:R-q}
\log (\frac{R_2}{R_1}) = (0.3893 \pm 0.0010)log q + (0.0001 \pm 0.0005)
\end{equation}

As shown in Figure \ref{Fig:rel}e,f, all 27 systems are in good agreement with the theoretical fits.

E) Based on the light curve solutions, the TIC 48219491, TIC 231718985, and TIC 322580598 systems have mass ratios of $q = 0.112$, $1/q = 0.109$, and $1/q = 0.129$, respectively. These systems are considered extremely low mass ratio contact binaries. We computed the ratio of spin to orbital angular momentum for these targets, using Equation \ref{eq:JJ}:

\begin{equation}\label{eq:JJ}
\frac{J_{\rm spin}}{J_{\rm orb}} = \frac{1+q}{q} \Big[ (k_1 r_1)^2 + (k_2 r_2)^2 q \Big],
\end{equation}
where $q$ is the mass ratio, $r_1$ and $r_2$ are the fractional radii of the primary and secondary, and $k_1$ and $k_2$ are the respective gyration radii. Using the mass ratios and radii derived in this study, we calculated the spin-to-orbital angular momentum ratio, by adopting $k_{1,2} = 0.06$ (\citealt{2004A&A...424..919C}). The calculated spin-to-orbital angular momentum ratios are 0.013(1) for TIC 48219491, 0.012(1) for TIC 231718985, and 0.015(1) for TIC 322580598. According to Darwin's instability criterion (\citealt{2015A&A...574A..39D}), since these values are much smaller than the theoretical limit of $\frac{1}{3}$, the spin angular momentum is negligible compared to the orbital angular momentum, indicating that the orbits of all three systems are stable.

F) The evolution of W UMa-type binary systems can be traced by examining the processes that cause a star to fill its Roche lobe. These processes are largely governed by changes in nuclear structure and angular momentum, both of which are strongly dependent on stellar mass (\citealt{1989MNRAS.237..447H}). Knowledge of the components' initial masses is essential for understanding the formation and subsequent evolution of W UMa-type binaries. The initial masses of the primary ($M_{1i}$) and secondary ($M_{2i}$) stars of target systems were estimated using the method described by the \cite{2013MNRAS.430.2029Y} study.

Estimation of the secondary component’s initial mass was carried out using the relation given in Equation \ref{Mi2} from the study by \cite{2013MNRAS.430.2029Y}.

\begin{equation}\label{Mi2}
M_{2i}=M_2+\Delta M=M_2+2.50(M_L-M_2-0.07)^{0.64},
\end{equation}
where $M_2$ is the current mass of the secondary, and $M_L$ is derived from the mass-luminosity relationship (Equation \ref{M_L}):

\begin{equation}\label{M_L}
M_L=\left(\frac{L_2}{1.49}\right)^{\frac{1}{4.216}}.
\end{equation}

Estimates of the primary component’s initial mass were obtained through the relation presented in Equation \ref{Mi1}:

\begin{equation}\label{Mi1}
M_{1i}=M_1-(\Delta M-M_{\text{lost}})=M_1-\Delta M(1-\gamma).
\end{equation}

The quantity $M_{\text{lost}}$ refers to the mass expelled from the system, and $\gamma$ is the ratio of $M_{\text{lost}}$ to $\Delta M$ (Equation \ref{Mlost}):

\begin{equation}\label{Mlost}
M_{\text{lost}}=\gamma \times \Delta M.
\end{equation}

A value of $\gamma=0.664$ was adopted following the findings reported by \cite{2013MNRAS.430.2029Y}. Table \ref{Tab:Mi} summarizes the outcomes, showing the initial masses of both components along with the estimated mass lost in each target system.

Table \ref{Tab:Mi} presents the initial masses of the primary components in our targets, which span 0.7–1.0 $M_{\odot}$. Stars within this mass range are likely to undergo relatively rapid angular momentum loss (\citealt{2004ARep...48..219T}). Our calculations show that the initial masses of the secondary components range from 0.8 to 2.4 $M_{\odot}$. The secondary stars have internal structures that differ substantially from those of normal main-sequence single-stars, due to their initial masses being higher than their current masses (\citealt{2013MNRAS.430.2029Y}). The difference between the current and initial masses indicates significant mass loss in most systems, and our results for all targets are consistent with those reported by \cite{2013MNRAS.430.2029Y}.

\begin{table*}
\renewcommand\arraystretch{1.2}
\caption{Initial masses of primary and secondary components and estimated mass loss for the target systems.}
\centering
\begin{center}
\footnotesize
\begin{tabular}{c c c c| c c c c}
\hline
TIC	&	$M_{1i}$($M_{\odot}$)	&	$M_{2i}$($M_{\odot}$)	&	$M_{\text{lost}}$($M_{\odot}$)	&	TIC	&	$M_{1i}$($M_{\odot}$)	&	$M_{2i}$($M_{\odot}$)	&	$M_{\text{lost}}$($M_{\odot}$)	\\
\hline
5800141	&	0.84(38)	&	1.51(24)	&	0.81(15)	&	243212894	&	0.76(36)	&	1.35(40)	&	0.62(25)	\\
23460412	&	0.78(34)	&	1.49(45)	&	0.64(28)	&	269943198	&	0.83(37)	&	1.50(33)	&	0.73(21)	\\
41740747	&	0.80(32)	&	1.66(18)	&	0.97(11)	&	276348408	&	0.98(33)	&	2.39(17)	&	1.41(11)	\\
48219491	&	0.94(32)	&	1.87(16)	&	1.13(10)	&	280651790	&	0.86(33)	&	1.66(27)	&	0.88(17)	\\
80428640	&	0.78(22)	&	0.93(36)	&	0.22(12)	&	287602202	&	0.75(35)	&	1.30(52)	&	0.46(20)	\\
82710288	&	0.81(34)	&	1.69(30)	&	0.87(19)	&	293775345	&	0.76(30)	&	1.69(30)	&	0.87(18)	\\
84445315	&	0.80(36)	&	1.18(53)	&	0.40(21)	&	296861174	&	0.67(37)	&	1.62(25)	&	0.89(16)	\\
89667365	&	0.84(39)	&	1.77(26)	&	0.95(16)	&	298708524	&	0.78(37)	&	0.75(29)	&	0.19(9)	\\
101040463	&	0.81(30)	&	1.59(19)	&	0.91(12)	&	322580598	&	0.89(34)	&	1.47(18)	&	0.86(11)	\\
138759051	&	0.88(31)	&	1.49(20)	&	0.84(12)	&	374271988	&	0.78(27)	&	1.61(26)	&	0.86(16)	\\
158546470	&	0.68(32)	&	1.39(54)	&	0.50(11)	&	386768115	&	0.80(31)	&	0.77(33)	&	0.04(4)	\\
184565222	&	0.82(32)	&	0.97(34)	&	0.15(6)	&	458718652	&	0.73(35)	&	1.54(34)	&	0.76(21)	\\
192854410	&	0.67(34)	&	1.27(47)	&	0.55(20)	&	468265910	&	0.83(37)	&	1.12(46)	&	0.36(13)	\\
231718985	&	0.97(32)	&	1.80(15)	&	1.08(9)	&		&		&		&		\\
\hline
\end{tabular}
\end{center}
\label{Tab:Mi}
\end{table*}

%%%%%%%%%%%%%%%%%%%%%%%%%%%%%%%%%%%%%%%%%%%%%%%%%%
\vspace{0.6cm}
\section*{Acknowledgments}
This manuscript was prepared within the framework of a workshop program conducted over more than six months in 2025 for selected talented and research-oriented participants from across Iran. This educational-research program was conducted by the Yazd Branch of the National Organization for Development of Exceptional Talents (SAMPAD).
We used data from the European Space Agency mission Gaia(\url{http://www.cosmos.esa.int/gaia}). In this work, we utilize observations obtained by the TESS mission, supported through NASA's Explorer Program.

%%%%%%%%%%%%%%%%%%%%%%%%%%%%%%%%%%%%%%%%%%%%%%%%%%%%
\vspace{0.6cm}
\bibliography{References}{}
\bibliographystyle{aasjournal}

\end{document}